\documentclass[12 pt,twoside,aps,prd,amsmath,amssymb,
tightenlines,showpacs,showkeys]{revtex4}
\usepackage{mathptmx}
\usepackage{bm}
\usepackage{graphicx}
{\hspace*{\fill}{\protect\small {\bf Bijan Saha}} \hspace*{\fill} }
{\hspace*{\fill} {\protect\small {\bf Non-minimally coupled
nonlinear spinor field in FRW cosmology}} \hspace*{\fill} }
\usepackage[colorlinks]{hyperref}

\newcommand {\G}{\Gamma}

\newcommand {\bg}{\bar \gamma}

\newcommand {\bp}{\bar \psi}

\begin{document}
\title{Non-minimally coupled
nonlinear spinor field in FRW cosmology}
\author{Bijan Saha}
\affiliation{Laboratory of Information Technologies\\
Joint Institute for Nuclear Research, Dubna\\
141980 Dubna, Moscow region, Russia\\ and\\
Institute of Physical Research and Technologies\\
People's Friendship University of Russia\\
Moscow, Russia} \email{bijan@jinr.ru}
\homepage{http://spinor.bijansaha.ru}

\hskip 1 cm

\begin{abstract}
Within the scope of a FRW cosmological model we have studied the
role of spinor field in the evolution of the Universe when it is
non-minimally coupled to the gravitational one. We have considered a
few types of nonlinearity. It was found that if the spinor field
nonlinearity describes an ordinary matter such as radiation, the
presence of non-minimality becomes essential and leads to the rapid
expansion of the Universe, whereas if the spinor field nonlinearity
describes a dark energy, the evolution of the Universe is dominated
by it and the difference between the minimal and non-minimally
coupled cases become almost indistinguishable.
\end{abstract}



\maketitle

\bigskip

\section{Introduction}

The discovery and further confirmation of the accelerated expansion
of the Universe led to reconsider the existing theories of
cosmology. One of the straight forward ways was to introduce some
additional component into the right hand side of the Einstein
equations with negative pressure which would work as repulsive force
thus giving rise to the accelerated mode of expansion. A number of
models were proposed by different authors. Model exploiting the
spinor field was one of them. For more than two decades spinor field
is being widely used in cosmology mainly thanks to its specific
behavior in presence of gravitational field. In a number of papers
the authors have shown that the nonlinear spinor field can give rise
to regular solutions as well as explain the late-time accelerated
mode of expansion of the Universe
\cite{Saha2001PRD,Saha2006PRD,Saha2009aECAA,ELKO,kremer,Saha2018ECAA}.
But most of those papers considered the minimal coupling of spinor
and gravitational field. It should be noted that along with  the
dark energy models many authors suggested the modification of the
Einstein equations itself. Scalar tensor theory cite{Brans-Dicke},
theory with non-minimal coupling, $F(R)$ theory \cite{Starobin},
$F(R,\,T)$ theory with $T$ being the trace of energy-momentum tensor
(EMT) \cite{Harko}, $F(T)$ theory with $T$ being the torsion
\cite{Li}, $f(G)$ theory are the few to name. The motivation behind
this research was to study the influence of spinor field in the
evolution of the universe when it is non-minimally coupled to the
gravitational field. Since spinor field is more sensitive to the
gravitational field than the scalar one, in our view it may give
rise to some unexpected results. Recently, Carloni {\it et al}
\cite{Astro-Phys/1811.10300} has considered non-minimally coupled
spinor field with the gravitational one. Non-minimally coupled
spinor and gravitational fields within the scope of Bianchi type-I
metric was studied in \cite{SpinBInm}. In this report we plan to
continue that study for an isotropic and homogeneous space-time
given by a FRW metric.

\section{Basic equations}

We consider the action in the form

\begin{equation}
S = \int \sqrt{-g} \left[\left(\kappa_1 + \lambda_1 S\right) R +
L_{\rm sp} \right] d \Omega, \quad \kappa_1 = \frac{1}{2 \kappa}.
\label{action}
\end{equation}
where $S = \bp\psi$ is a scalar constructed from spinor fields,
$\lambda_1$ is the coupling constant. Here $\kappa$ is the
Einstein's constant defined as $\kappa = 8 \pi G$, with $G$ being
the Newton's gravitational constant. The spinor field Lagrangian
takes the form

\begin{equation}
L_{\rm sp} = \frac{\imath}{2} \left[\bp \gamma^{\mu} \nabla_{\mu}
\psi- \nabla_{\mu} \bar \psi \gamma^{\mu} \psi \right] - m \bp \psi
- \lambda F (S). \label{lspin}
\end{equation}
Note that in general the nonlinear term $F$ may be the arbitrary
function of invariant $K$ which takes one of the following
expressions: $\{I,\,J,\,I+J,\,I-J\}$. Here $I = \bp \psi$ and $J =
\imath \bp \bg^5 \psi$. Here $m$ is the spinor mass. $\lambda$ is
the self coupling constant that can be positive or negative. Here
$\nabla_{\mu}$ is the covariant derivative of the spinor field
\begin{equation}
\nabla_\mu \psi = \partial_\mu \psi - \Gamma_\mu \psi, \quad
\nabla_\mu \bp = \partial_\mu \bp + \bp \Gamma_\mu. \label{covder}
\end{equation}
Here $\Gamma_\mu$ is the spinor affine connection.

Variation with respect to metric functions give \cite{SpinBInm}
\begin{equation}
R_\mu^\nu - \frac{1}{2} \delta_\mu^\nu R = \frac{1}{\left(\kappa_1 +
\lambda_1 S \right)}\left[T_\mu^\nu + \lambda_1
\left(g^{\nu\tau}\nabla_\mu \nabla_\tau - \delta_\mu^\nu
\Box\right)S\right]. \label{EE}
\end{equation}
where $T_\mu^\nu$ is the energy-momentum tensor of the spinor field.
The corresponding equations for spinor field we find varying the
action with respect to $\psi$ and $\bp$ \cite{SpinBInm}

\begin{subequations}
\label{speq}
\begin{eqnarray}
\imath\gamma^\mu \nabla_\mu \psi - m \psi - \lambda F_S \psi + \lambda_1 R \psi &=&0, \label{speq1} \\
\imath \nabla_\mu \bp \gamma^\mu +  m \bp + \lambda F_S \bp  -
\lambda_1 R \bp &=& 0. \label{speq2}
\end{eqnarray}
\end{subequations}
From \eqref{speq} one finds that $L_{\rm sp} = SF_S - F.$

We consider the isotropic FRW space-time is given by
\begin{equation}
ds^2 = dt^2 - a^{2}\left(\,dx_1^{2} + \,dx_2^{2} +\,dx_3^2\right),
\label{frw}
\end{equation}
with the scale factor $a$ is the functions of time only.

For the metric \eqref{frw} we choose the tetrad such that they have
the following nontrivial components:

\begin{equation}
e_0^{(0)} = 1, \quad e_i^{(i)} = a,  \quad i = 1,\,2,\,3.
\label{tetradfrw}
\end{equation}

From

\begin{eqnarray}
\Gamma_\mu &=& \frac{1}{8}\left[\partial_\mu \gamma_\alpha,
\gamma^\alpha\right] - \frac{1}{8}
\Gamma^{\beta}_{\mu\alpha}\left[\gamma_\beta, \gamma^\alpha\right].
\label{SPAC}
\end{eqnarray}
where $\left[a,b\right] = a b - b a$ one finds the following
expressions for spinor affine connections:
\begin{equation}
\G_0 = 0, \quad \G_1 = \frac{\dot a}{2} \bg^1 \bg^0, \quad \G_2 =
\frac{\dot a}{2} \bg^2 \bg^0, \quad \G_3 = \frac{\dot a}{2} \bg^3
\bg^0. \label{sac123frw}
\end{equation}
In \eqref{SPAC} $\gamma_\beta = e_\beta^{(b)} \bg_b$ and
$\gamma^\alpha = e^\alpha_{(a)} \bg^a$ are the Dirac matrices in
curve space-time and $e^\alpha_{(a)}$ and $e_\beta^{(b)}$ are the
tetrad vectors.

We consider the case when the spinor field depends on $t$ only. The
spinor field equations in this case read

\begin{subequations}
\label{speqfrw}
\begin{eqnarray}
\imath\bg^0\left( \dot \psi + \frac{3}{2}\frac{\dot a}{a}
\psi\right)
- m \psi - \lambda F_S \psi + \lambda_1 R \psi &=&0, \label{speq1frw} \\
\imath \left( \dot \bp  + \frac{3}{2}  \frac{\dot a}{a} \bp\right)
\bg^0 + m \bp + \lambda F_S \bp  - \lambda_1 R \bp &=& 0,
\label{speq2frw}
\end{eqnarray}
\end{subequations}
where we denote $F_S = dF/dS.$ From \eqref{speqfrw} one easily finds

\begin{equation}
S = \frac{C_0}{a^3}, \quad C_0 = {\rm Const.} \label{Sa}
\end{equation}

The energy-momentum tensor of the spinor field
\begin{equation}
T_{\mu}^{\,\,\,\rho}=\frac{\imath}{4} g^{\rho\nu} \biggl(\bp
\gamma_\mu \nabla_\nu \psi + \bp \gamma_\nu \nabla_\mu \psi -
\nabla_\mu \bar \psi \gamma_\nu \psi - \nabla_\nu \bp \gamma_\mu
\psi \biggr) \,- \delta_{\mu}^{\rho} L_{\rm sp}, \label{temsp}
\end{equation}
in our case gives the following nontrivial components
\cite{SpinBInm}
\begin{subequations}
\label{Ttotfrw}
\begin{eqnarray}
T_0^0 & = &  m S + \lambda F(S),\label{Ttot00frw}\\
T_1^1 & = & T_2^2 = T_3^3 = \lambda\left( F(S) - S
F_S\right).\label{Ttotiifrw}
\end{eqnarray}
\end{subequations}

Taking into account that in our case,  $\Box S = \ddot S +
3\frac{\dot a}{a} \dot S$, in view of
\begin{subequations}
\label{nablaS}
\begin{eqnarray}
\nabla_\mu \nabla_\nu S &=& \nabla_\mu \partial_\nu S = \partial_\mu
\partial_\nu S - \Gamma^\alpha_{\mu \nu} \partial_\alpha S,
\label{nabla1}\\
\Box S &=& g^{\alpha \beta} \nabla_\alpha \nabla_\beta S = g^{\alpha
\beta} \left(\partial_\alpha
\partial_\beta S - \Gamma^\tau_{\alpha \beta} \partial_\tau
S\right), \label{nablaS2}
\end{eqnarray}
\end{subequations}

for the metric \eqref{frw} from \eqref{EE} we find
\begin{subequations}
\label{EEComp}
\begin{eqnarray}
2\frac{\ddot a}{a} + \frac{{\dot a}^2}{a^2}&=&
\frac{1}{\left(\kappa_1 + \lambda_1 S \right)} \left[\lambda \left(
F(S) - S F_S\right) -
\lambda_1 \ddot S - 2 \lambda_1 \frac{\dot a}{a} \dot S\right],\label{11frw}\\
3\frac{{\dot a}^2}{a^2}&=& \frac{1}{\left(\kappa_1 + \lambda_1 S
\right)} \left[ \left(m S + \lambda F(S)\right) - 3\lambda_1
\frac{\dot a}{a} \dot S\right]. \label{00frw}
\end{eqnarray}
\end{subequations}

In a recent paper \cite{SpinBInm} it was shown that if instead of
ordinary scalar we deal with $S = \bp\psi$ as component by
component, then for the second derivative in \eqref{nabla1} we get
an additional term, namely $\bp \Gamma_\nu
\partial_\mu \psi - \partial_\mu \bp \Gamma_\nu \psi$. In our case
$\psi$ is a function of $t$, moreover $\G_0 = 0$ and $\G_i = (\dot
a/2) \bg^i \bg^0$. On account of that we can write $\bp \Gamma_\nu
\partial_\mu \psi - \partial_\mu \bp \Gamma_\nu \psi = \frac{\dot a}{2} \left(\bp
\bg^i \bg^0 \dot \psi - \dot \bp \bg^i \bg^0 \psi\right).$ Further
multiplying \eqref{speq1frw} by $\bp \bg^i$ from the left and
\eqref{speq2frw} by $\bg^i \psi$ from the right and adding them we
find $ \left(\bp \bg^i \bg^0 \dot \psi - \dot \bp \bg^i \bg^0
\psi\right) \equiv 0.$ Thus in this case we can deal with $S$ as an
ordinary scalar.

In view of \eqref{Sa} from \eqref{00frw} we find

\begin{equation}
\dot a = \sqrt{\frac{\left(m S + \lambda F\right)}{3\left(\kappa_1
-2 \lambda_1 S \right)}}\, a = \sqrt{\frac{\left(m C_0 + \lambda a^3
F\right)}{3\left(\kappa_1 a^3 -2 \lambda_1 C_0 \right)}} \, a.
\label{a2asS}
\end{equation}

Further from \eqref{Sa} we find that $\ddot S = - 3\frac{\ddot a}{a}
S + 12 \frac{{\dot a}^2}{a^2}S$. Then on account of \eqref{a2asS} we
rewrite \eqref{11frw} as

\begin{equation}
\ddot a = \left[\frac{\lambda \left(F - S
F_S\right)}{\left(2\kappa_1 - \lambda_1 S\right)} -
\frac{\left(\kappa_1 + 7 \lambda_1 S\right)\left(mS + \lambda F
\right)}{3\left(2\kappa_1 - \lambda_1 S\right)\left(\kappa_1 -
2\lambda_1 S\right)}\right] a, \label{afinal}
\end{equation}

in view of \eqref{Sa} which can be written as

\begin{equation}
\ddot a = \left[\frac{\lambda \left(a^3 F - C_0
F_S\right)}{\left(2\kappa_1 a^3 - \lambda_1 C_0\right)} -
\frac{\left(\kappa_1 a^3 + 7 \lambda_1 C_0 \right)\left(m C_0 +
\lambda a^3 F \right)}{3\left(2\kappa_1 a^3 - \lambda_1
C_0\right)\left(\kappa_1 a^3 - 2\lambda_1 C_0\right)}\right] a.
\label{afinal0}
\end{equation}

\section{Numerical analysis}

In what follows we solve this equation numerically. For simplicity
we set $m = 1$, $\kappa_1 = 1$ and $C_0 = 1$. We consider two cases,
one with non-minimal coupling, another with minimal coupling, so
that the role of non-minimal coupling becomes clear.

{\bf Case 1} In this case we consider the non-minimal coupling with
nonlinear term (plotted in solid blue line) setting $\lambda_1 =
1,\, \lambda = 1$.

{\bf Case 2} As a second case we consider nonlinear spinor field
with minimal coupling setting  $\lambda_1 = 0,\, \lambda = 1$
(plotted in dot red line).

The initial value $a(0)$ was chosen in such a way that the initial value
of $\dot a(0)$  that was determined from \eqref{a2asS} remains real.
As it was mentioned earlier, the
nonlinear spinor field can simulate different types of dark energy.
Here we consider different types of nonlinearity and compare the
results for there different cases.

{\bf Dust}

\begin{figure}[ht]
\centering
\includegraphics[height=70mm]{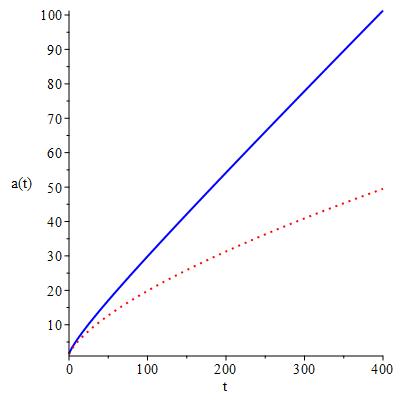} \\ \vskip 1 cm
\caption{Plot of scale factor $a$ with the Universe filled with
dust. Blue solid line stands for non-minimal coupling, while the red
dot line corresponds to minimal coupling.} \label{FRW-Dust}
\end{figure}

Let us begin with linear spinor field. Setting $\lambda = 0$ from
 \eqref{Ttotfrw} we find $T_0^0 = m S$ and $T_1^1 = T_2^2 = T_3^3 = 0$.
 It means the linear spinor field behaves like dust. In Fig.~\ref{FRW-Dust}
 the behavior or scale factor $a$ is plotted for non-minimal and minimal
 coupling. As one sees, non-minimal coupling in this case leads to
 th e rapid expansion of the Universe.

{\bf Radiation}

\begin{figure}[ht]
\centering
\includegraphics[height=70mm]{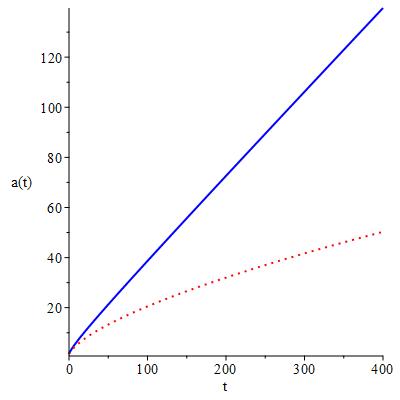} \\ \vskip 1 cm
\caption{Plot of scale factor $a$ with the Universe filled with
radiation. Blue solid line stands for non-minimal coupling, while
the red dot line corresponds to minimal coupling.}
\label{FRW-Radiation}
\end{figure}

Let us first consider the case when the Universe is filled with
radiation. In this case the spinor field nonlinearity is given by
\cite{Saha2018ECAA}
\begin{equation}
F = S^{1 + W}, \quad W =  1/3. \label{rad}
\end{equation}
The corresponding solution is given in Fig.~\ref{FRW-Radiation}.
Here the blue solid line stand for non-minimal coupling with
nonlinear spinor field, and red dot line stands for minimal coupling
with nonlinear spinor field. Like in the previous case here too we
see that the non-minimal coupling plays significant role in the
evolution of the Universe and leads to its rapid expansion.

{\bf Quintessence}

\begin{figure}[ht]
\centering
\includegraphics[height=70mm]{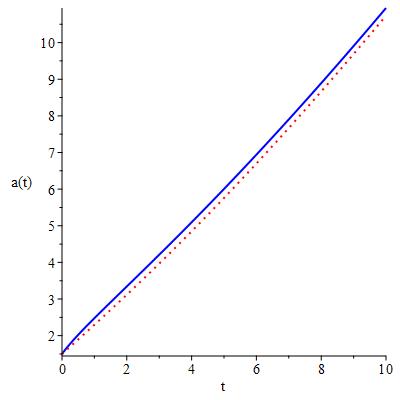} \\ \vskip 1 cm
\caption{Plot of scale factor $a$ with the Universe filled with
quintessence. Blue solid line stands for non-minimal coupling, while
the red dot line corresponds to minimal coupling.} \label{FRW-qunit}
\end{figure}

Let us consider the spinor field nonlinearity which is responsible for quintessence.
In this case the spinor field nonlinearity can be given by  \cite{Saha2018ECAA}
\begin{equation}
F = S^{1 + W}, \quad  W < - 1/3. \label{quint}
\end{equation}
Let us set  $W = -1/2$. The solution to the equation \eqref{afinal0}
is plotted in the Fig.~\ref {FRW-qunit}. Here we see that the
nonlinear term plays the key role in the evolution of the Universe.
The presence of non-minimality is hardly distinguishable.

{\bf Chaplygin Gas}

\begin{figure}[ht]
\centering
\includegraphics[height=70mm]{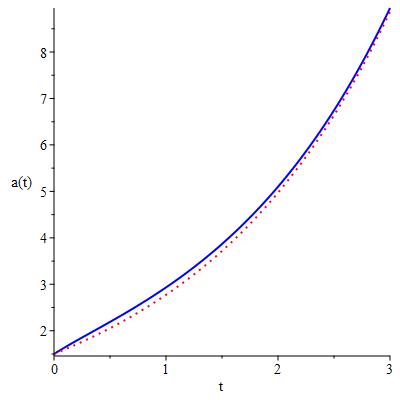} \\ \vskip 1 cm
\caption{Plot of scale factor $a$ with the Universe filled with
Chaplygin gas. Blue solid line stands for non-minimal coupling,
while the red dot line corresponds to minimal coupling.}
\label{FRW-Chap}
\end{figure}

Another choice of spinor field nonlinearity could be the one that
describes a Chaplygin gas. As it was shown in \cite{Saha2018ECAA}
spinor field nonlinearity in this case takes the form
\begin{equation}
F = \left(A + S^{\left(1 +
\alpha\right)}\right)^{1/\left(1 + \alpha\right)}, \label{Chap}
\end{equation}
with $ A> 0$ and $0 < \alpha \le 1$. Inserting it into
\eqref{afinal0} and setting $A = 1$ and $\alpha = 0.5$ we have
solved the equation is question numerically. The result is
illustrated in Fig.~ \ref{FRW-Chap}. As in case of quintessence,
here too the prime role belongs to the spinor field nonlinearity.

{\bf Modified Quintessence}

\begin{figure}[ht]
\centering
\includegraphics[height=70mm]{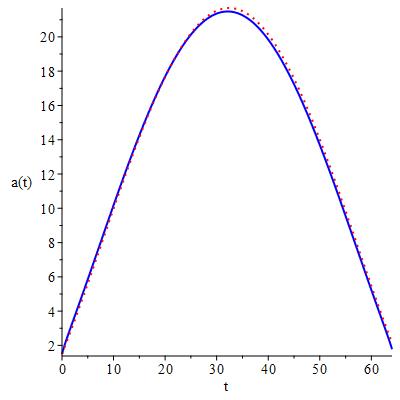} \\ \vskip 1 cm
\caption{Plot of scale factor $a$ with the Universe filled with
modified quintessence. Blue solid line stands for non-minimal
coupling, while the red dot line corresponds to minimal coupling.}
\label{FRW-ModQuint}
\end{figure}

The discovery of late time acceleration gives rise a number of
problems. One of the problems is the eternal acceleration. To avoid
this a modified quintessence was proposed. In this case the spinor
field nonlinearity takes the form \cite{Saha2018ECAA}
\begin{equation}
F = S^{1 + W} + \frac{W}{1+W} \varepsilon_{\rm cr}, \label{Modquint}
\end{equation}
where  $\varepsilon_{\rm cr}$ is come constant. We set  $ W = - 1/2$
and $\varepsilon_{\rm cr} = 0.01$ Then the solution to the equation
\eqref{afinal0} takes the from drawn in Fig.~\ref{FRW-ModQuint}. We
again see that the evolution of the Universe is dominated by the
dark energy given by the spinor field nonlinearity.

{\bf Modified Chaplygin Gas}

We also consider the case when the dark energy is the combination of
quintessence and Chaplygin gas. In this case the spinor field
nonlinearity takes the form \cite{Saha2018ECAA}
\begin{equation}
F = \left( \frac{A}{1+W} + S^{\left(1 +
W\right)\left(1 + \alpha\right)}\right)^{1/\left(1 + \alpha\right)}, \label{ModChap}
\end{equation}
with $ W < - 1/3$ and $A > 0$. We have taken $W = -1/2$ and $\alpha
= 2.$ The corresponding solution is illustrated in the
Fig.~\ref{ModChap}. Like other previous cases spinor field
nonlinearity plays principal role in the evolution of the Universe.

\begin{figure}[ht]
\centering
\includegraphics[height=70mm]{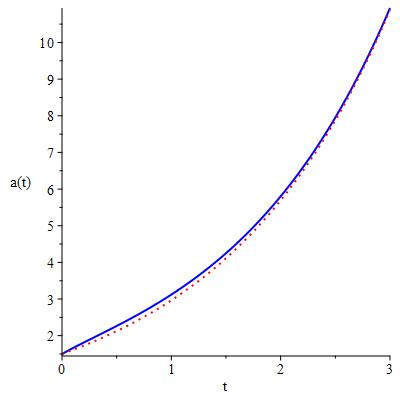} \\ \vskip 1 cm
\caption{Plot of scale factor $a$ with the Universe filled with
modified Chaplygin gas. Blue solid line stands for non-minimal
coupling, while the red dot line corresponds to minimal coupling.}
\label{FRW-ModChap}
\end{figure}

\section{conclusion}

Since in a FRW Universe the non-diagonal components of the
energy-momentum tensor of the spinor field do not exist the spinor
field does not impose any additional restriction on the geometry of
the Universe as it takes place for the anisotropic cosmological
models. This is true for both cases with minimal and non-minimal
coupling. If the spinor field nonlinearity behaves like an ordinary
matter, e.g., radiation, the presence of non-minimality becomes
significant and in this case the non-minimal coupling leads to the
rapid rapid expansion of the Universe, whereas if the spinor field
nonlinearity describes a dark energy, the evolution of the Universe
is totally dominated by it and the presence of non-minimality
remains almost unnoticeable.

\vskip 7mm

\noindent {\bf Acknowledgments}\\
This work is supported in part by a joint Romanian-JINR, Dubna
Research Project, Order no.396/27.05.2019 p-71.

\end{document}